# The Generalized PT-Symmetric Sinh-Gordon Potential Solvable within Quantum Hamilton-Jacobi Formalism

Özlem Yeşiltaş[a], S. Bilge Ocak[b,*]

[a] Gazi University, Faculty of Arts and Sciences, Department of Physics, 06500, Teknikokullar, Ankara, Turkiye

[b] Sarayköy Nuclear Research and Training Center
Kazan, Ankara, Turkiye.

**Abstract**

The generalized Sinh-Gordon potential is solved within quantum Hamiltonian Jacobi approach in the framework of $PT$ symmetry. The quasi exact solutions of energy eigenvalues and eigenfunctions of the generalized Sinh-Gordon potential are found for $n = 0, 1$ states.

[*] Corresponding Author: semamuzo@yahoo.com

# 1. Introduction

The discovery of new class of physically significant spectral problems, called quasi-exactly solvable (QES) models, has attracted much attention [1-3]. Several methods [4-5] for the generation of QES model have been worked out. These are the models for which a part of the bound state energy spectrum and corresponding wave functions can be obtained exactly. These models have been constructed and studied extensively by means of Lie algebraic approach. In order that a part of the spectrum can be obtained exactly, the parameters appearing in the relation must satisfy a condition known as the condition for quasi-exact solvability.

Within the Quantum Hamiltonian Jacobi approach (QHJ) [6-7], it has been found to be an elegant and the simple method to determine the energy spectrum of exactly solvable models in quantum mechanics. The advantage of this method is that it is possible to determine the energy eigen-values without having to solve for the eigen-functions. In this formalism, a quantum analog of classical action angle variables is introduced [6-7]. The quantization condition represents well known results on the number nodes of the wave function, translated in terms of logarithmic derivative, also is called quantum momentum function (QMF) [3, 6-7]. The equation satisfied by the QMF is a non-linear differential equation, called quantum Hamilton-jacobi equation leads to two solutions. The application of QHJ to eigen-values has been explored in great detail in ref. [3, 6-9].

In recent years, $PT$-symmetric Hamiltonians have generated much interest in quantum mechanics [10-11]. Physical motivation for the $PT$- symmetric but non-Hermitian Hamiltonians have been emphasized by many authors [12-16]. Recently, Mostafazadeh [16, 17] has introduced a different concept which is known as the class of pseudo-Hermitian Hamiltonians, and argued that the basic structure responsible for the particular spectral properties of these Hamiltonians is their pseudo-Hermiticity [17]. Following these detailed works, non-Hermitian Hamiltonians with real or complex spectra have been analyzed by using various numerical and analytical techniques [18-22]. In the applications, Ranjani and her collaborators applied the QHJ formalism, to Hamiltonians with Khare-Mandal potential and Scarf potential, characterized by discrete parity and time reversal $PT$ symmetries [3].

The aim of the present work is to calculate the energy eigenvalues and the corresponding eigenfunctions of the Sinh-Gordon potential field which is used in coherent spin states [23]. Originally the form of the potential is a hyperbolic one given as

$$V_0(x) = \frac{B^2}{4\gamma}\sinh^2 x - B(S+\frac{1}{2})\cosh x \qquad (1)$$

We study general form of this hyperbolic effective potential which is called as Sinh-Gordon potential and is also in PT-symmetric form where the potential parameters $V_1$ and $V_2$ are real [23],

$$V(x) = V_1 Sinh^2 \alpha x + V_2 Cosh \alpha x \qquad (2)$$

in a Hamiltonian system generating quasi-exactly-solvable problems.

In this letter, we first give a brief description of the QHJ formalism. The Sinh-Gordon potential is studied in one dimension. In each of these cases the condition of quasi-exactly solvability is derived within the QHJ approach. The energy and eigenfunctions of the generalized hyperbolic effective potential are obtained by using the QES condition.

## 2. Quantum Hamiltonian-Jacobi Formalism

In the quantum Hamilton-Jacobi (QHJ) formalism, the logarithmic derivative of the wave function $\psi(x)$ is given by [3, 6-7] :

$$p = -i\hbar \frac{d}{dx}(\ln \psi) \qquad (3)$$

which is called as quantum momentum function QMF. QMF plays an important role since it is stated analogous to the classical momentum function as, $p = \frac{dS}{dx}$ where $S$ is the Hamilton's characteristic function and it is related to the wave function by $\psi(x) = \exp\left(\frac{iS}{\hbar}\right)$. Substituting $\psi(x)$ in terms of $S$ in the Schrödinger equation $H\psi = E\psi$ and using the potential relation in Eq.(2), the following expression is obtained:

$$p^2 - i\hbar p' = 2m[E - V(x)]. \qquad (4)$$

Leacock and Padgett [6-7] proposed the quantization condition for the bound states in order to find the eigenvalues. Quantum action variable is defined as QMF and exact quantization condition for the bound states of a real potential is given as [3, 6-7]

$$\frac{1}{2\pi} \oint_C p\, dx = n\hbar \qquad (5)$$

where, C is the contour enclosing the n moving poles in the complex domain [3, 6-7].

## 3. The Sinh-Gordon Potential

The Sinh-Gordon potential which is given by Eq.(2) is $PT$ symmetric if $V_1$ and $V_2$ are real. It is worth noting that, parity operation is given by $p \to -p$, $x \to -x$ in the Hamiltonian whereas the time reversal remains same as the conventional $p \to -p$, $x \to x$ and $i \to -i$. The QHJ equation with $\hbar = 2m = 1$, can be introduced as:

$$p^2 - \frac{dp}{dx} - [E - V_1 \sinh^2 \alpha x - V_2 \cosh \alpha x] = 0 \quad (6)$$

In order to transform Eq. (6) into a rational form, one can change the variable $y = \cosh \alpha x$ in Eq.(6) and obtain the following relation:

$$p^2(y) - i\alpha \sqrt{y^2 - 1}\, p'(y) - [E - V_1(y^2 - 1) - V_2 y] = 0 \quad (7)$$

The coefficient of the $\frac{dp}{dx} \neq 1$ in Eq.(7). Thus, in order to bring it to the Riccati type equation form, $p$ can be defined as;

$$p = -i\alpha \sqrt{y^2 - 1}\, \phi \quad (8)$$

And and another transformation is given as below,

$$\phi = \chi - \frac{y}{2(y^2 - 1)}. \quad (9)$$

Using Eqs.(8) and (9) in Eq.(7), one can obtain a Riccati type equation as

$$\chi' + \chi^2 + \frac{y^2 + 2}{4(y^2 - 1)^2} + \frac{1}{\alpha^2(y^2 - 1)}\left[E - V_1 y^2 - V_2 y + V_1\right] = 0 \quad (10)$$

Assume that $\chi$ has a finite number of moving poles in the complex $y$ plane. In addition to, $\chi$ has not only moving poles, but also fixed poles at $y = \pm 1$ and it is bounded at $y = \infty$. Then, $\chi$ can be written in the form of,

$$\chi = \frac{b_1}{y-1} + \frac{b_1'}{y+1} + \frac{P_n'(y)}{P_n(y)} + C \qquad (11)$$

where $b_1$ and $b_1'$ are the residues at $y = \pm 1$. $P_n(y)$ is a polynomial of degree $n$ and $C$ is a constant due to the Liouville's theorem (for more details, see [3]). At $y = 1$, $\chi$ is expanded in a Laurent series as:

$$\chi = \frac{b_1}{y-1} + a_0 + a_1(y-1) + \ldots \qquad (12)$$

Substituting Eq.(12) in Eq. (10), one obtains the values of residue $b_1$ as:

$$b_1 = \frac{1}{4}, \frac{3}{4} \qquad (13)$$

and also $a_0 = \pm \frac{\sqrt{V_1}}{\alpha}$ is found. For large $y$, $\chi \to \pm \frac{\sqrt{V_1}}{\alpha}$ which are the values of $C$. If the same procedure is applied to $\chi$ for $y = -1$

$$\chi = \frac{b_1'}{y+1} + a_0' + a_1'(y+1) \qquad (14)$$

One can obtain the residue at $y = -1$ as

$$b_1' = \frac{1}{4}, \frac{3}{4} \qquad (15)$$

In order to the discuss the behaviour of $\chi$ at infinity, one expands $\chi$ as:

$$\chi = a_0 + \frac{\lambda}{y} + \frac{\lambda_1}{y^2} + \ldots \qquad (16)$$

Substitution of Eq.(16) in (10) gives

$$\lambda = \frac{V_2}{2\sqrt{V_1}\alpha}, -\frac{V_2}{2\sqrt{V_1}\alpha} \qquad (17)$$

Behaviour of $\chi$ approaches to $\dfrac{b_1 + b_1' + n}{y}$ for large $y$. Hence one can obtain [3],

$$b_1 + b_1' + n = \lambda \qquad (18)$$

It can be seen that Eq.(18) is positive. Therefore, one can choose positive value of $\lambda$ which is $\dfrac{V_2}{2\sqrt{V_1}\,\alpha}$ and it leads to choose $a_0 = C = -\dfrac{\sqrt{V_1}}{\alpha}$ because of the physical solutions for the wavefunction. In order to obtain a QES condition, one must take all possible combinations of the residues $b_1, b_1'$. These combinations with a constraint are given in table 3.1 as:

**Table3.1.** The QES condition and the $n$ for each $b_1, b_1'$.

| set | $b_1$ | $b_2$ | $n = \lambda - b_1 - b_1'$ | Condition on $M = \dfrac{V_2}{2\sqrt{V_1}\,\alpha}$ | | QES condition |
|---|---|---|---|---|---|---|
| 1 | $\dfrac{1}{4}$ | $\dfrac{1}{4}$ | $\dfrac{M-1}{2}$ | $M$ : odd, | $M \geq 1$ | $M = 2n+1$ |
| 2 | $\dfrac{3}{4}$ | $\dfrac{3}{4}$ | $\dfrac{M-3}{2}$ | $M$ : odd, | $M \geq 3$ | $M = 2n+3$ |
| 3 | $\dfrac{3}{4}$ | $\dfrac{1}{4}$ | $\dfrac{M-2}{2}$ | $M$ : even, | $M \geq 2$ | $M = 2n+2$ |
| 4 | $\dfrac{1}{4}$ | $\dfrac{3}{4}$ | $\dfrac{M-2}{2}$ | $M$ : even, | $M \geq 2$ | $M = 2n+2$ |

As it is seen from the table3.1 that sets 1 and 2 are valid if and only if $M$ is odd, sets 3 and 4 are valid if and only if $M$ is even. When it comes to the form of the wavefunction, using $\psi(x) = \exp(i \int p\, dx)$ and writing $p$ in terms of $\chi$, it is obtained as

$$\psi(y) = \exp \int \left( \frac{b_1}{y-1} + \frac{b_1'}{y+1} + \frac{P_n'}{P_n} - \frac{\sqrt{V_1}}{\sqrt{\alpha}} - \frac{y}{2(y^2-1)} \right) dy \qquad (19)$$

Substitute Eq.(19) in (10) to obtain,

$$\frac{P_n''}{P_n} + \frac{2P_n'}{P_n}\left( \frac{b_1}{y-1} + \frac{b_1'}{y+1} + \frac{\sqrt{V_1}}{\alpha} \right) + \frac{b_1^2 - b_1}{(y-1)^2} + \frac{b_1'^2 - b_1'}{(y+1)^2} +$$

$$\frac{2\sqrt{V_1}}{\alpha}\left( \frac{b_1}{y-1} + \frac{b_1'}{y+1} \right) + \frac{2b_1 b_1'}{y^2-1} + \frac{y^2+2}{4(y^2-1)^2} - \frac{V_1}{\alpha^2} + \qquad (20)$$

$$\frac{1}{\alpha^2(y^2-1)}\left[ E - V_1(y^2-1) - V_2 y \right] = 0.$$

Using Eq.(20), one can find the energy eigenvalues and corresponding wavefunctions for $M = 3$ and $M = 2$ cases. For $M = 3$ state, it can be seen that $n = 1$ and the $P_n(y)$ is a first order polynomial given as $P_1 = \gamma y + \beta$. The results are illustrated in table3.2 and table3.3 respectively.

**Table3.2.** When $M = 3$ and for set 1-2, the energy eigenvalues and wave-functions.

| $M = 3$ | $b_1, b_1'$ | Energy, $E_n$ | Wave-function $\psi_n$ |
|---|---|---|---|
| Set 1 | $b_1 = b_1' = \dfrac{1}{4}$, $n = 1$ | $E_1 = -\dfrac{\alpha^2}{4} + \alpha\sqrt{V_1}$ | $\psi_1(x) = e^{-\dfrac{\sqrt{V_1}}{\alpha}\cosh\alpha x}(\gamma\cosh\alpha x + \beta)$ |
| Set 2 | $b_1 = b_1' = \dfrac{3}{4}$, $n = 0$ | $E_0 = -\alpha^2$ | $\psi_0(x) = e^{-\dfrac{\sqrt{V_1}}{\alpha}\cosh\alpha x}(\sinh\alpha x)$ |

where $\beta = 1$ and $\gamma = -\dfrac{12\sqrt{V_1}}{-3 + 4\sqrt{V_1}}$ are found for $c\,\mathrm{sgn}(\sqrt{V_1}) \succ 0$ and $\gamma = -\dfrac{4\sqrt{V_1}}{-3 + 4\sqrt{V_1}}$ is found for $c\,\mathrm{sgn}(\sqrt{V_1}) \prec 0$. Similarly for $M = 2$ case, the table3.3 is given as

**Table3.3.** When $M = 2$ and for set 3-4, the energy eigenvalues and wave-functions.

| $M = 2$ | $b_1, b_1'$ | Energy, $E_n$ | Wave-function $\psi_n$ |
|---|---|---|---|
| Set 3 | $b_1 = \dfrac{1}{4}, b_1' = \dfrac{3}{4}$, $n = 0$ | $E_0 = -\dfrac{\alpha^2}{4} - \alpha\sqrt{V_1}$ | $\psi_0(x) = e^{-\dfrac{\sqrt{V_1}}{\alpha}\cosh\alpha x}(\cosh\alpha x + 1)$ |
| Set 4 | $b_1 = \dfrac{3}{4}, b_1' = \dfrac{1}{4}$, $n = 0$ | $E_0 = -\dfrac{\alpha^2}{4} - \alpha\sqrt{V_1}$ | $\psi_0(x) = e^{-\dfrac{\sqrt{V_1}}{\alpha}\cosh\alpha x}(\cosh\alpha x - 1)$ |

From table3.2 and table 3.3, the energy eigenvalues are obtained for the odd and even values of $\dfrac{V_2}{2\sqrt{V_1}\alpha}$ for Sinh-Gordon potential.

If the complex forms of the potential given in Eq.(2) are written for $\alpha = 2$ as below,

$$V(x) = V_1 \sinh^2 2x + iV_2 \cosh 2x \qquad (21)$$

$$V(x) = iV_1 \sinh^2 2x + V_2 \cosh 2x \qquad (22)$$

the solutions of QHJ equation for these complex potentials can be discussed. Eq.(21) is $PT$ symmetric under parity reflections as $x \to \frac{i\pi}{2} - x$, $i \to -i$ and Eq.(22) is non-$PT$ symmetric. If we take Eq.(21) into consideration, values of $\lambda$ can be found by following the same procedure as,

$$\lambda = \pm \frac{iV_2}{4\sqrt{V_1}} \qquad (23)$$

which also means the values of $M$ in the table3.1 are complex. The same procedure is valid for the potential relation in Eq.(22), too. Hence, these conditions don't lead to physical solutions (for the wavefunctions). But, in ref.[3] the potential has the form given below

$$V(x) = -(\varsigma \cosh 2x - iM)^2 \qquad (24)$$

which is $PT$ symmetric under parity reflections $x \to \frac{i\pi}{2} - x$, $i \to -i$ and leads to real energy spectrum [3].

**Conclusions**

In this letter quantum Hamilton-Jacobi formalism is briefly given and the generalized Sinh-Gordon potential is solved within quantum Hamiltonian Jacobi approach in the framework of $PT$ symmetry. We have obtained the quasi exact solutions of eigenvalues and eigenfunctions of the potential for $n = 0, 1, 2$ states. The general Sinh-Gordon potential is applied to the coherent spin states. The eigenvalues and eigenfunctions comply with the results of ref.[23], for $S = 0, 1/2$ ($\gamma = 1$) spin states if the potential parameters are chosen appropriately.

Interesting features of quantum expectation theory for $PT$-violating potentials may be affected by changing from complex to real systems. Finally, we have pointed out that our exact results of the Sinh-Gordon potential may increase the number of applications in the study of various nonlinear potential applications in quantum systems.

**Acknowledgements**

This research was partially supported by the Scientific and Technological Research Council of Turkiye.

# References

[1] A. Khare, U. Sukhatme, J.Math.Phys. 40 (1999) 5473-5494.

[2] F. Cooper, A. Khare, U. Sukhatme, Phys. Rep. 251 (1995) 267.

[3] K. G. Geojo, S. S. Ranjani and A. K. Kapoor, J. of Phys. A: Math and Gen.,36 (2003) 309 ; S.S. Ranjani et al, Mod. Phys. Lett. A 19 (2004) 1457-1468; S.S. Ranjani et al, Int. J. of Modern Physics A, 20 (2005) 4067-4077.

[4] Carl M. Bender, Stefan Boettcher, J.Phys. A 31 (1998) L273-L277.

[5] B. Bagchi, P. Gorain et al, Czech. J. Phys. 54 (2004) 1019-1026.

[6] R. A. Leacock and M. J. Padgett, Phys. Rev. Lett., 50 (1983) 3.

[7] R. A. Leacock and M. J. Padgett, Phys. Rev. D 28 (1983) 2491.

[8] R. S. Bhalla, A.K. Kapoor and P.K. Panigrahi, Am. J. Phys. 65 (1997) 1187.

[9] R. S. Bhalla, A.K. Kapoor and P.K. Panigrahi, Mod. Phys. Lett. A, 12 (1997) 295.

[10] C. M. Bender, J. of Math. Phys., 37 (1996) 6-11 .

[11] C. M. Bender and S. Boettcher, Phys. Rev. Lett., 80 (1998) 5243; C. M. Bender and S. Boettcher and P. N. Meisenger, J. Math. Phys., 40 (1999), 2201; C. M. Bender, G. V. Dunne and P. N. Meisenger, Phys. Lett. A 252 (1999) 272.

[12] Ö.Yeşiltaş, M. Şimşek, R. Sever, C. Tezcan, Physica Scripta, 67 (2003) 472.

[13] M. Aktaş, R. Sever, Mod.Phys.Lett. A 19 (2004) 2871-2877.

[14] M. Znojil, J.Phys., A 32 (1999) 7419-7428.

[15] M. Znojil, J. Phys. A:Math. and Gen., 36 (2003) 7639-7648.

[16] A. Mostafazadeh, J.of Math. Phys., 43 (2002) 205-214 .

[17] A. Mostafazadeh, A. Batal, J.Phys. A 37 (2004) 11645-11680; A. Mostafazadeh, J. Phys. A: Math. Gen., 38 (2005) 3213-3234; A. Mostafazadeh, J. Math. Phys. 46, (2005) 1-15.

[18] F. Cannata et all, Phys. Lett. A246 (1998) 219.

[19] A. Khare, B.P. Mandal, Phys. Lett.A 272 (2000) 53.

[20] B. Bagchi, C. Quesne, Phys. Lett. A 273 (2000) 285.

[21] Z. Ahmed, Phys. Lett. A 282 (2000) 343.

[22] C. M. Bender et all, Phys. Lett. A 291 (2000) 197.

[23] O. B. Zaslavskii and V. V. Ulyanov, Sov. Pyhs. JETP 60 (1984) 991.